\documentclass[11pt,twoside]{article}
\usepackage{asp2010}
\usepackage{color}

\resetcounters

\begin{document}

\title{Resonant Compton Upscattering in High Field Neutron Stars} 
\author{Peter L. Gonthier$^1$, Matthew T. Eiles$^1$, Zorawar Wadiasingh$^2$,\\ and Matthew G. Baring$^2$
\affil{$^1$Department of Physics, Hope College, 27 Graves Place, \\ Holland, Michigan 49423}
\affil{$^2$Department of Physics and Astronomy, Rice University, 6100 Main Street, Houston, Texas 77005-1827}}

\begin{abstract}
The extremely efficient process of resonant Compton upscattering by relativistic electrons in high magnetic fields is believed to be a leading emission mechanism of high field pulsars and magnetars in the production of intense X-ray radiation. New analytic developments for the Compton scattering cross section using Sokolov \& Ternov (S\&T) states with spin-dependent resonant widths are presented. These new results display significant numerical departures from both the traditional cross section using spin-averaged widths, and also from the spin-dependent cross section that employs the Johnson \& Lippmann (J\&L) basis states, thereby motivating the astrophysical deployment of this updated resonant Compton formulation. Useful approximate analytic forms for the cross section in the cyclotron resonance are developed for S\&T basis states. These calculations are applied to an inner magnetospheric model of the hard X-ray spectral tails in magnetars, recently detected by RXTE and INTEGRAL. Relativistic electrons cool rapidly near the stellar surface in the presence of intense baths of thermal X-ray photons. We present resonant Compton cooling rates for electrons, and the resulting photon spectra 
at various magnetospheric locales, for magnetic fields above the quantum critical value. These demonstrate how this scattering mechanism has the potential to produce the characteristically flat spectral tails observed in magnetars.
\end{abstract}

Recent observations reveal the presence of hard X-ray tails in the spectra measured by INTEGRAL, RXTE, XMM and ASCA in Anomalous X-ray Pulsars ( \citet{K04,K06}; \citet{Hart8a,Hart8b}) as well as in Soft Gamma-ray Repeaters (\citet{Mereghetti}; \citet{Molkov}; \citet{Gotz}; \citet{Rea09}).  Resonant Compton scattering (RCS) is an efficient mechanism that may interpret these observations. RCS has been studied by several groups (\citet{LG06}; \citet{Rea08}; \citet{BH07}; \citet{NTZ08}; \citet*{BWG11} ) within this context.

We develop compact, analytical expressions to correctly describe resonant, magnetic inverse Compton scattering from relativistic electrons in the neutron star magnetospheres. The objective is to explore the potential for such upscattering to generate the high-energy tails observed in magnetars.  These expressions are developed using the \citep*{ST68} (S\&T)  basis states, which several groups have shown (\citet*{BGH05,G93,MP83} ) to be the formally correct choice of basis states.  We develop the description of Compton scattering with a resonance that is intrinsically spin-dependent within the context of the commonly used J\&L as well as S\&T electron basis states (Gonthier et al. 2012, in preparation).  As indicated in \cite*{BGH05}, J\&L states do not behave appropriately, mixing spin states under Lorentz boosts along the field resulting in the lifetimes of the excited states that are incorrectly characterized.  Given the broad use within the literature of J\&L states in resonant Compton scattering, we proceed in a parallel analysis of both S\&T and J\&L basis states.

The correct formalism of resonant Compton scattering using the spin-dependent resonance width developed with proper S\&T electron wave functions displays significant differences with respect to the traditionally-used cross section with the average width, and lesser differences relative to the resonant cross section employing the spin-dependent width using the J\&L electron wave functions.
These were used to compute lepton cooling rates for arbitrary magnetospheric locales for
interactions with X-rays emitted from the surface.  The rates generally 
indicate that cooling in the resonance should dramatically lower the electron Lorentz factor as it traverses 
field lines to moderate altitudes.  We also present  
inverse Compton spectra for relativistic electrons of fixed Lorentz factor, 
specifically for different observer viewing perspectives relative 
to the dipolar field axis.  The spectra were integrated over entire electron transits of closed magnetic field loops of various 
maximum altitudes, but coincident with the plane defined by the field axis and the observer direction.  
These spectra displayed an array of forms, with a general trend of being harder when the observer views 
from a hemisphere opposite to the one where the electron originates.
Preliminary indications are that spectra close to those observed in
10-150 keV tails in magnetars can be generated by electrons that possess
Lorentz factors of around 10--50.  These strongly encourage the development of a full radiation 
emission model that encompasses electron cooling, to explain the AXP and SGR observations;
such a more complete model defines our upcoming research on this problem.

\acknowledgements We are also grateful for the generous support of the National Science Foundation (grants AST-1009731 and REU PHY/DMR-1004811), and the NASA Astrophysics Theory Program through grants  NNX09AQ71G and NNX10AC59A.

\bibliography{GonthierPoster}

\begin{thebibliography}{}
\expandafter\ifx\csname natexlab\endcsname\relax\def\natexlab#1{#1}\fi
\expandafter\ifx\csname url\endcsname\relax
  \def\url#1{\texttt{#1}}\fi
\expandafter\ifx\csname urlprefix\endcsname\relax\def\urlprefix{URL }\fi
\providecommand{\eprint}[2][]{\url{#2}}

\bibitem[{Baring et~al.(2005)Baring, Gonthier, \& Harding}]{BGH05}
Baring, M.~G., Gonthier, P.~L., \& Harding, A.~K. 2005, ApJ, 630, 430

\bibitem[{Baring \& Harding(2007)}]{BH07}
Baring, M.~G., \& Harding, A.~K. 2007, Ap\&SS, 308, 109

\bibitem[{Baring et~al.(2011)Baring, Wadiasingh, \& Gonthier}]{BWG11}
Baring, M.~G., Wadiasingh, Z., \& Gonthier, P.~L. 2011, ApJ, 733, 61

\bibitem[{den Hartog et~al.(2008{\natexlab{a}})den Hartog, Kuiper, \&
  Hermsen}]{Hart8a}
den Hartog, P.~R., Kuiper, L., \& Hermsen, W. 2008{\natexlab{a}}, A\&A, 489,
  263

\bibitem[{den Hartog et~al.(2008{\natexlab{b}})den Hartog, Kuiper, Hermsen,
  Kaspi, Dib, Knödlseder, \& Gavriil}]{Hart8b}
den Hartog, P.~R., Kuiper, L., Hermsen, W., Kaspi, V.~M., Dib, R., Knödlseder,
  J., \& Gavriil, F.~P. 2008{\natexlab{b}}, A\&A, 489, 245

\bibitem[{G{\"o}tz et~al.(2006)G{\"o}tz, Mereghetti, Tiengo, \&
  Esposito}]{Gotz}
G{\"o}tz, D., Mereghetti, S., Tiengo, A., \& Esposito, P. 2006, A\&A, 449, L31

\bibitem[{{Graziani}(1993)}]{G93}
{Graziani}, C. 1993, \apj, 412, 351

\bibitem[{Kuiper et~al.(2006)Kuiper, Hermsen, den Hartog, \& Collmar}]{K06}
Kuiper, L., Hermsen, W., den Hartog, P.~R., \& Collmar, W. 2006, ApJ, 645, 556

\bibitem[{Kuiper et~al.(2004)Kuiper, Hermsen, \& Mende{\'{z}}}]{K04}
Kuiper, L., Hermsen, W., \& Mende{\'{z}}, M. 2004, ApJ, 613, 1173

\bibitem[{{Lyutikov} \& {Gavriil}(2006)}]{LG06}
{Lyutikov}, M., \& {Gavriil}, F.~P. 2006, \mnras, 368, 690.
  \eprint{arXiv:astro-ph/0507557}

\bibitem[{{Melrose} \& {Parle}(1983)}]{MP83}
{Melrose}, D.~B., \& {Parle}, A.~J. 1983, Aust. J. Phys., 36, 755

\bibitem[{Mereghetti et~al.(2005)Mereghetti, Götz, Mirabel, \&
  Hurley}]{Mereghetti}
Mereghetti, S., Götz, D., Mirabel, I.~F., \& Hurley, K. 2005, A\&A, 433, L9

\bibitem[{Molkov et~al.(2005)Molkov, Hurley, Sunyaev, Shtykovsky, Revnivtsev,
  \& Kouveliotou}]{Molkov}
Molkov, S., Hurley, K., Sunyaev, R., Shtykovsky, P., Revnivtsev, M., \&
  Kouveliotou, C. 2005, A\&A, 433, L13

\bibitem[{{Nobili} et~al.(2008){Nobili}, {Turolla}, \& {Zane}}]{NTZ08}
{Nobili}, L., {Turolla}, R., \& {Zane}, S. 2008, \mnras, 389, 989.
  \eprint{0806.3714}

\bibitem[{{Rea} et~al.(2008){Rea}, {Zane}, {Turolla}, {Lyutikov}, \&
  {G{\"o}tz}}]{Rea08}
{Rea}, N., {Zane}, S., {Turolla}, R., {Lyutikov}, M., \& {G{\"o}tz}, D. 2008,
  \apj, 686, 1245. \eprint{0802.1923}

\bibitem[{Rea et~al.(2009)}]{Rea09}
Rea, N., et~al. 2009, MNRAS, 396, 2419

\bibitem[{Sokolov \& Ternov(1968)}]{ST68}
Sokolov, A.~A., \& Ternov, I.~M. 1968, Synchrotron Radiation (Oxford: Pergamon)

\end{thebibliography}

\end{document}